\documentclass[11pt,twocolumn]{article}

\usepackage{graphicx}
\usepackage{dcolumn}
\usepackage{amsmath,amsfonts}
\usepackage{bm}
\usepackage[numbers]{natbib}
\usepackage[latin1]{inputenc}
\usepackage{booktabs}
\usepackage{abstract}

\setcitestyle{square}
\title{Fast computation of loudness using a deep neural network}

\author{Josef Schlittenlacher, Richard, E. Turner, Brian C. J. Moore}

\date{University of Cambridge\\%
    js2251,ret26,bcjm@cam.ac.uk\\[2ex]%
    \today}

\begin{document}
\maketitle

\begin{abstract}
The present paper introduces a deep neural network (DNN) for predicting the instantaneous loudness of a sound from its time waveform. The DNN was trained using the output of a more complex model, called the Cambridge loudness model. While a modern PC can perform a few hundred loudness computations per second using the Cambridge loudness model, it can perform more than 100,000 per second using the DNN, allowing real-time calculation of loudness. The root-mean-square deviation between the predictions of instantaneous loudness level using the two models was less than 0.5 phon for unseen types of sound. We think that the general approach of simulating a complex perceptual model by a much faster DNN can be applied to other perceptual models to make them run in real time.
\end{abstract}

\section{\label{sec:intro}Introduction}

Accurate models for predicting perceptual attributes of sound (such as loudness) from their physical characteristics can have high computational cost, often making it hard or impossible to run them in real time. For example in the auditory domain, a good  model needs to estimate the input to the auditory nerve, which is determined by the excitation pattern in the cochlea. The excitation at a given place in the cochlea is a non-linear function of the sound's momentary spectrum and depends not only on frequency, but also on level and interactions between adjacent frequencies. In addition, transformations are needed between various scales, some of which are neither linear nor logarithmic.

One of the most advanced loudness models \cite{ISO5322,ISO5323} (see \cite{Moore2014} for an overview, or \cite{Moore2018} for the most recent update), which we call the Cambridge loudness model, uses the time waveform of a sound as the input and calculates three quantities: (1) Instantaneous loudness, which is the momentary loudness calculated from a given frame of the sound and which is assumed not to be available for conscious perception; (2)~Short-term loudness, which is the loudness of a short segment of the sound, such as a single syllable in a sentence; (3) Long-term loudness, which the overall loudness impression of a longer segment of sound, such as a whole sentence. The most computationally intensive step is step (1). In the model, the instantaneous loudness is updated every 1 ms, a rate that is necessary to accommodate the temporal resolution of the auditory system \cite{Zwicker1974}. However, on a modern PC it is only possible to calculate instantaneous loudness a few hundred times per second. This means that the Cambridge loudness model cannot be run in real time. Furthermore, it would be desirable to have a computation speed that is much faster than real time, for example when calculating the time-varying loudness of long recordings of sound (sometimes durations of days or weeks are needed to evaluate environmental noise), or when estimating individual model parameters in an active-learning test \cite{Schlittenlacher2018}, where as many evaluations as possible within an acceptable inter-trial interval of less than about two seconds are desired.

For this reason we developed a deep neural network (DNN) for predicting instantaneous loudness from a given input spectrum, using instantaneous loudness calculated from the Cambridge loudness model as a reference for training. Predicted values were expressed as loudness level in phon; the loudness level of a given sound is defined as the sound pressure level of an equally loud 1-kHz tone presented in free field with frontal incidence. After training,  the root-mean-square (RMS) difference between the loudness level predicted by the DNN and by the Cambridge loudness model was less than 0.5 phon for sounds of unseen categories. This error is somewhat below the just noticeable difference. Our implementation in Keras/TensorFlow can calculate instantaneous loudness more than 100,000 times per second on a CPU (i7 6700k).

\section{Model}

Apart from accuracy, computation speed was the main consideration when designing the DNN. The Cambridge loudness model estimates the short-term spectrum in each frame using six Fourier Transforms in parallel, each being used to estimate the spectrum in a limited frequency region. For the DNN, the input was a simplified spectrum with 61 components covering the frequency range up to 8 kHz (constant-width bins up to 200 Hz, nine bins per octave above 200 Hz). The limit of 8 kHz was chosen due to the limited sampling rate of the training material.

The output was a single loudness level estimate in phon. This scale was chosen because of its similarity to the input scale, which was measured in decibels. The two scales range roughly from 0 to 100 (between the detection threshold and the point at which sounds become uncomfortably loud), and the just noticeable difference in loudness is roughly constant on these scales. This made it easier for the DNN to develop a mapping from input to output without the need for the scale transformations and summations across frequency that are required in the Cambridge loudness model. Furthermore, the use of the phon scale as output made it possible to use simple ReLU activations \cite{Nair2010}. When operating an auditory DNN on other scales, for example the waveform, the combination of sigmoid and hyperbolic tangent can give better results \cite{Oord2016}.

The DNN was a multilayer perceptron (MLP) that consisted of an input layer with 61 units, three hidden layers with 150 units each, and a single output unit with linear activation. It was optimized with regard to the mean square difference between the DNN and the Cambridge loudness model. The Adam optimizer \cite{Kingma2014} was used with its default parameters. All weights were initialized randomly.

Alternative architectures were also evaluated. Convolutional neural networks did not achieve the same accuracy. A likely reason for this is that the input scale (logarithmic frequency) differs from the ERB-number scale, which is a perceptually relevant frequency scale based on estimates of the bandwidths of the auditory filters \cite{Glasberg1990}, and thus filters for low and high frequencies need considerably different shapes.

The training data consisted of three different types of sounds. First, 500,000 spectra were calculated from the LibriSpeech corpus \cite{Panayotov2015}, using the ``clean" development set. The sounds were scaled to have an overall RMS level of 60 dB\,SPL. Spectra were calculated every 35 ms (560 samples) using a 1024-point discrete Fourier Transform (DFT). Second, about 700,000 pure tones with levels ranging from -15 to 110 dB SPL and various levels of background noise were generated. Each component of the background noise was at least 10 dB lower than the level of the pure tone. Third, about 500,000 spectra of band-limited noises and noises with spectral notches were generated. They had various overall levels, bandwidths, notch widths and spectral gradients. The DNN was first trained for 220 epochs, then for a further 780, and then for a further 4000.

\section{Experiments}

\begin{figure}[htb]
    \includegraphics[width=1\columnwidth]{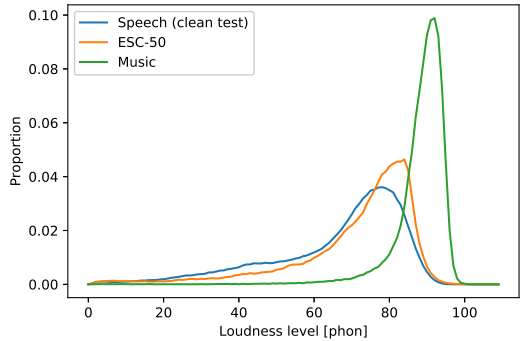}
    \caption{Distributions of loudness levels predicted by the DNN, shown as the proportions that fall within each 1-phon wide bin. The majority of loudness levels lie close to the RMS value (phon are somewhat higher than levels in dB SPL because of spectral loudness summation), although some predicted loudness levels fall well below the RMS level for both speech and the ESC set. }
    \label{fig:dist}
\end{figure}

Loudness was predicted for two further sets of data from the LibriSpeech corpus, ``clean" test and ``other" test. Each of them consisted of 500,000 spectra and they were calibrated to have an RMS level of 60 dB\,SPL. Loudness was also predicted for 250,000 spectra from the ESC corpus \cite{Piczak2015}. This corpus contains 50 categories of environmental sounds, for example rain, animals, aircraft, keyboard typing or a washing machine. The sounds were again scaled to have an RMS level of 60 dB\,SPL. Furthermore, loudness was predicted for 100,000 spectra from 20 popular songs of the 1960s, which were scaled to have an RMS level of 70~dB\,SPL. The predicted loudness distributions are shown in Figure \ref{fig:dist}. For the speech sets, only results for the ``clean" test are shown since the distributions were virtually the same for the ``other" test.
All loudness calculations of the Cambridge loudness model were based on the 1024-point DFT, while predictions of the DNN were based on the simplified 61-point input spectrum, which in turn was obtained from the DFT spectrum.

Table \ref{tab:error} shows the RMS difference in phon between the predictions of the Cambridge loudness model and the predictions of the DNN, which is referred to as the error. The RMS error for clean speech of 0.27 phon after 1000 epochs is virtually the same as the training error. The RMS error is less than 0.5 phon for ``other" speech, which according to the developers of the corpus is somewhat more noisy, the music, and most notably for the environmental sounds. The value of 0.5 phon is similar to or below the just noticeable difference for loudness, i.e. most predictions deviate by an amount that is less than the amount needed for a human listener to distinguish them.

\begin{table} [htbp]
\caption{\label{tab:error} {\small RMS error in phon between predictions of the DNN and the Cambridge loudness model }} 
\vspace{2mm}
\centerline{
\begin{tabular}{lrrr}
 & \multicolumn{3}{c}{Epochs} \\
 & 220 & 1000 & 5000 \\
\toprule
LibriSpech ``clean" test & 0.35 & 0.27 & 0.28 \\
LibriSpech ``other" test & 0.55 & 0.45 & 0.47 \\
ESC-50 & 0.56 & 0.45 & 0.47 \\
1960s songs & 0.38 & 0.35 & 0.31 \\
\bottomrule
\end{tabular}}
\end{table}

\begin{figure}[htb]
    \includegraphics[width=1\columnwidth]{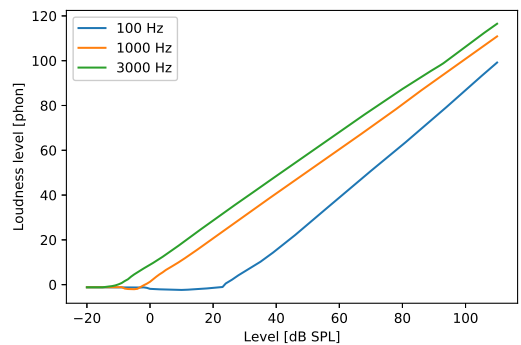}
    \caption{Loudness level of 3-kHz, 1-kHz and 100-Hz tones as a function of sound level, as predicted by the DNN after 1000 epochs.}
    \label{fig:1}
\end{figure}

Figure \ref{fig:1} shows the predicted loudness level of pure tones in quiet as a function of input sound level. The lowest loudness level predicted by the Cambridge loudness model was limited to 0 phon, since the threshold in quiet corresponds to about 2 phon for a normal-hearing listener. The loudness level is systematically higher for the 3-kHz tone than for the 1-kHz tone because 3 kHz is near to the resonant frequency of the ear canal. The threshold is about 20 dB higher at 100 Hz than it is at 1 kHz, but the difference in loudness loudness decreases with increasing level. All these predicted effects correspond well to loudness judgments obtained from human listeners.

\begin{figure}[htb]
    \includegraphics[width=1\columnwidth]{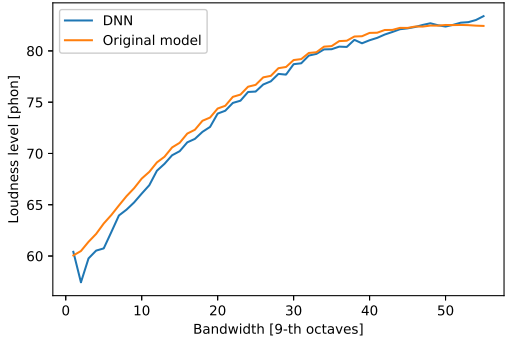}
    \caption{Loudness level of a bandpass-filtered pink noise with an overall level of 60 dB SPL and centered at 1 kHz as a function of its bandwidth.}
    \label{fig:2}
\end{figure}

Figure \ref{fig:2} shows the loudness level of bandpass filtered pink noise centered at 1 kHz, plotted as a function of bandwidth, as predicted by the Cambridge loudness model and the DNN. The predictions of the DNN are a little below of the Cambridge loudness model, especially for small bandwidths. These deviations are probably due to the fact that the DNN does not sum the loudness density across frequency at any stage, but rather performs a regression from the 61 input levels to the output loudness levels. It is of interest, however, that the results predicted by the DNN are more consistent with recent psychophysical results \cite{Hots2014}. Note that the sounds used for figures \ref{fig:1} and \ref{fig:2} were presented to the DNN during training. The predictions for these sounds are shown because the effects of frequency and spectral summation are fundamental aspects of loudness.

\section{Conclusions}

The predictions for the environmental sounds and music are remarkably accurate given that the DNN was trained using speech and synthetic sounds only. This suggests that the DNN generalizes well to real-world sounds. The predictions for music with slightly higher loudness levels showed that the DNN also works well for levels to which it has been exposed less frequently. Training using pure tones and noises ensured that the effects of level, frequency and spectral loudness summation would be represented adequately, and probably led to better generalization than training solely using speech. Using an adversarial example \cite{Szegedy2013}, it might be possible to find spectra for which predictions of the Cambridge loudness model and the DNN deviate more. We leave this for a future study and conclude for now that the DNN generalizes well to a variety of real-world sounds.

In summary, we developed and evaluated a DNN that was trained using the predictions of a computationally more expensive model, the Cambridge loudness model. The gain in computational speed was a factor of more than 100, enabling computation much faster than real-time, while predictions were almost the same. This allows real-time prediction of loudness with accuracy comparable to that for the Cambridge loudness model. The DNN would also be useful for the analysis of large amounts of pre-recorded data. The approach of using a DNN to approximate a perceptual model could readily be extended to searches for individual model parameters in efficient hearing tests. Another extension could be in devices like hearing aids and cochlear implants to allow hearing to be restored more nearly to normal.

\section*{Acknowledgments}

The work was supported by the Engineering and Physical Sciences Research Council (UK, grant number RG78536).

\bibliography{literature}%

\end{document}